\documentclass[useAMS]{mn2e}
\usepackage{lscape}
\usepackage{rotating}
\usepackage{amssymb}

\def\hi{\hbox{H{\sc i}}}
\def\micron{\hbox{$\mu$m}}

\def\msun{\hbox{M$_\odot$}}

\def\t4{\hbox{t$_{\rm 4}$}}

\def\cm3{\hbox{cm$^{-3}$}}

\input psfig.sty
\input epsf.sty
\usepackage{graphicx}
\usepackage{epsfig}
\title[A lack of gas in young LMC/SMC clusters]
{Constraining globular cluster formation through studies of young massive clusters - III. A lack of gas and dust in massive stellar clusters in the LMC and SMC}
\author[Bastian \& Strader]{N. Bastian$^1$ \& J. Strader$^2$ \\
$^{1}$ Astrophysics Research Institute, Liverpool John Moores University, 146 Brownlow Hill, Liverpool L3 5RF, UK\\
$^{2}$ Department of Physics and Astronomy, Michigan State University, East Lansing, MI 48824, USA \\
}
\date{Accepted. Received; in original form}
\pagerange{\pageref{firstpage}--\pageref{lastpage}}
\pubyear{2013}
\begin{document}
\maketitle
\label{firstpage}
\begin{abstract}
Scenarios that invoke multiple episodes of star formation within young globular clusters (GCs) to explain the observed chemical and photometric anomalies in GCs, require that clusters can retain the stellar ejecta of the stars within them and accrete large amounts of gas from their surroundings.  Hence, it should be possible to find young massive clusters in the local Universe that contain significant amounts ($>10$\%) of the cluster mass of gas and/or dust within them.  Recent theoretical studies have suggested that clusters in the Large Magellanic Cloud (LMC) with masses in excess of $10^4$\msun\, and ages between $30$ and $\sim300$~Myr, should contain such gas reservoirs.  We have searched for \hi\ gas within 12 LMC (and 1 SMC) clusters and also for dust using {\em Spitzer} 70\micron\ and 160\micron\ images.  No clusters were found to contain gas and/or dust.  While two of the clusters have \hi\ at the same (projected) position and velocity, the gas does not appear to be centred on the clusters, but rather part of nearby clouds or filaments, suggesting that the gas and cluster are not directly related.  This lack of gas ($<1$\% of the stellar mass) is in strong tension with model predictions, and may be due to higher stellar feedback than has been previously assumed or due to the assumptions used in the previous calculations.

\end{abstract}
\begin{keywords} 
\end{keywords}

\section{Introduction}
\label{sec:intro}

Stellar clusters have traditionally been thought of as simple stellar populations (SSPs), with all of the stars within a given cluster having the same abundance and age, within some small tolerance.  However, all old globular clusters (GCs) in the Galaxy that have been studied in the necessary detail show abundance anomalies not observed in field stars of the same metallicity. The most common of these is a Na-O anti-correlation (e.g., Carretta et al. 2009). Many GCs with precise photometry from the Hubble Space Telescope also show features  in the colour-magnitude diagram that are inconsistent with being SSPs, such as multiple sequences, in regions from the main sequence turnoff to the giant branch. The origin of these anomalies is still under debate.

One potential solution is to have multiple epochs of star formation within massive clusters, i.e., when the globular clusters were young ($<200-300$~Myr).  In this scenario, material processed by stars of an initial population in the form of stellar ejecta is mixed with a large amount of primordial (unenriched) gas, which then forms a second (or multiple) generations.  The stars that contribute the processed material have been suggested to be rapidly rotating massive stars (e.g., Decressin et al.~2007), interacting high mass binaries (e.g., de Mink et al.~2009) and Asymptotic Giant Branch (AGB) stars (e.g., D'Ercole et al.~2008), among other candidates.  If AGB stars are the contributor, which do not process material for the first $\sim30$~Myr of a cluster's life, the cluster must be able to retain large amounts of primordial gas or accrete new gas from its surroundings (e.g., Conroy \& Spergel 2011 - hereafter CS11).

CS11 have theoretically investigated the conditions necessary for a cluster to accrete gas from its surroundings.  There are two main mechanisms to bring gas into an existing cluster 1) through Bondi-Hoyle type accretion into the cluster and 2) if the cluster already has some fraction of its mass in gas, that internal gas can sweep up material as the cluster passes through the ISM.  In the second case, the cluster gas can also be stripped from the cluster due to ram pressure.  

CS11 have applied their calculations to clusters in the LMC, taking the observed current ISM density and cluster velocities within the galaxy.  The authors suggest that there is a minimum mass, $\sim10^4$\msun, above which clusters should be able to retain (and further accrete) gas from their surroundings.  These clusters may then undergo a second star formation epoch. 

Such secondary (or continuous or multiple-burst) have been suggested to have happened in a number of intermediate age ($1-2$~Gyr), massive ($>10^4\msun$) clusters in the LMC and SMC (e.g., Mackey \& Broby Nielsen~2007; Milone et al.~2009; Goudfrooij et al. 2011a,b).  These works have suggested (semi-)continuous star-forming episodes lasting for $200-500$~Myr based on extended main-sequence turn-offs within the clusters.  However, observations of massive ($10^4 - 10^5$\msun) clusters in the LMC with younger ages, where such age spreads should be readily apparent, show that these clusters are all consistent with a single burst of star-formation ($\Delta(age) < 10-40$~Myr - Bastian \& Silva-Villa~2013; Niederhofer et al.~2014).  Additionally, observations of $130$ Galactic and extragalactic clusters with masses between $10^4 - 10^8$\msun\ and ages between $10-1000$~Myr found no evidence of ongoing star-formation within any of the clusters (Bastian et al.~2013a).  If star-formation continued for $100-200$~Myr after an initial burst, at least 50\% of their sample should have shown evidence for ongoing star-formation, unless the initial mass function of stars was severely deficient in high mass stars ($>15$\msun).  These observations call into question whether clusters are able to retain/accrete the necessary gas and dust in order to form stars after the initial burst.

If the anomalies seen in the intermediate age LMC clusters (i.e., the extended main sequence turn-off - eMSTO) are due to (continuous) extended star-formation histories, then at any given time during the star-formation epoch, a significant amount of gas needs to be present.  The exact amount depends on when the cluster is observed during its extended formation.  Bastian et al.~(2013a) have estimated the amount of ongoing star-formation, as a function of the current mass of the clusters, expected in clusters with different star-formation histories.  If we adopt the continuous star-formation histories estimated by Goudfrooij et al.~(2011b), which are reasonably well approximated by Gaussians with dispersions of $200-500$~Myr, then the clusters would be expected to be presently forming $>7-10\%$ of their current mass (averaged over 7~Myr).  For 100\% star-formation efficiency, we then would expect $>700-1000$~\msun\ of gas within the clusters in our survey (M$_{\rm cluster,stars} > 10^4$\msun).

However, the amount of gas predicted to be in clusters by CS11 is not as straight-forward.  While the Bondi-Hoyle accretion of the gas is a direct function of the cluster mass and ISM density, most of the gas in their model is swept up from the surrounding ISM by the initial gas in the cluster (i.e. the cluster gas acts as a net as the cluster passes through the ISM).  CS11 adopt an initial gas mass of $10$\% of the mass of the cluster, which is effective in sweeping up the material, hence a lower limit predicted for the amount of gas present in our sample would be $10^3$~\msun, which would increase as the clusters accrete gas from their surroundings.

We note an importance difference between the CS11 models and the observations and interpretation presented by Mackey \& Broby Nielsen~(2007) and Goudfrooij et al.~(2011).  The CS11 models predict that gas stays within the cluster for the first 100-200~Myr, and then cools rapidly, allowing a second, discrete, burst of star-formation.  However, the observations suggest a smooth spread in the eMSTOs of the intermediate age clusters, which suggests (if interpreted as being caused by differences in age) a continuous star-formation episode lasting $\sim500$~Myr.  Additionally, D'Ercole, D'Antona, \& Vesperini~(2011) have argued against the CS11 scenario on the basis of the observed chemistry within GCs.  If AGB stars are the source of the enriched material, in order to produce the ``extreme" stars (in terms of their abundances), these stars need to be formed entirely from material processed through AGB stars, i.e. without any diluting pristine material.  The CS11 scenario results in a smooth mass accretion, whereas D'Ercole et al.~(2011) argue that a time variable accretion rate of the pristine material from the surroundings is required to fit the observations.

In the present work, we test the above predictions, that young massive clusters in the LMC (and SMC) have significant amounts of gas within them. We study a sample of clusters with masses $>10^4$\msun\ and ages between 25 and 315~Myr (see \S~\ref{sec:clusters}), i.e., those predicted to host a gas reservoir.  In our search, we are guided by the predictions of CS11, that the gas should be cool ($T \sim 100$~K), so should be bright in \hi\, and possibly in dust emission.  We use the publicly available ATCA+Parkes \hi\ surveys of the LMC (Staveley-Smith et al. 2003; Kim  et al.~2003) and SMC (Stanimirovi\'c et al. 1999) to search for \hi\ at the position and velocity of the clusters in our sample.  For the dust we use the {\em Spitzer SAGE Survey} (Meixner et al.~2006) and the {\em SAGE-SMC Survey} (Gordon et al.~2011) to search for emission at the location of the clusters, and we also search for optical colour gradients within the clusters, as the gas/dust is expected to be centrally concentrated in the cluster, so the inner portions are predicted to be redder than the outer parts.

We note that alternative theories have been put forward to explain the observed abundance anomalies in GCs and extended main sequence turn-offs in the intermediate age LMC/SMC clusters that do not invoke multiple (or continuous) epochs of star-formation.  For GCs this is the early disc accretion scenario (Bastian et al.~2013b) and for the extended main-sequence turn-offs, both stellar rotation (Bastian \& de Mink~2009) and interacting binary stars (Yang et al.~2011) have been suggested.


The paper is organised as follows, in \S~\ref{sec:obs} we describe the cluster sample as well as the observations used in the present work.  In \S~\ref{sec:results} we present our results on the search for gas/dust within the clusters and in \S~\ref{sec:discussion} we present our conclusions.


\section{Observations and Techniques}
\label{sec:obs}

\subsection{Cluster catalogue}
\label{sec:clusters}

We selected our sample of clusters from the McLaughlin \& van der Marel~(2005) catalogue, searching for clusters in the LMC/SMC with ages between $10-500$~Myr and masses in excess of $10^4$\msun.  The masses were obtained from profile fitting as well as by comparison of the cluster luminosities with the mass-to-light ratio of simple stellar population models of the appropriate age.  The clusters and their ages and masses are given in Table~\ref{tab:clusters}, mostly from McLaughln \& van der Marel~(2005), however, NGC~1850, NGC~1856 and NGC~1866 have updated values taken from Niederhofer et al.~(2014) and Bastian \& Silva-Villa (2013).  None of the clusters in the current sample have been suggested to have significant age spreads within their stellar populations.  However, the sample clusters were predicted by CS11 to host significant amounts of gas within them and have the potential to form further stellar generations.


\subsection{H{\sc i} data}
\label{sec:hi}

Kim et al.~(2003), Staveley-Smith et al. (2003), and Stanimirovic et al.~(1999) published \hi\ maps of the LMC and SMC that combined single-dish (Parkes) and multi-dish (ATCA) for sensitivity to gas at a range of scales. The 1--1.6\arcmin\ resolution of these surveys is comparable to the typical sizes of massive clusters in the Magellanic Clouds, and so is a reasonable match (modulo the large uncertainties in model predictions) to the scales on which we would expect to find gas if it were present.

For each star cluster in the survey we extracted \hi\ emission spectra at the location of the cluster\footnote{http://www.atnf.csiro.au/research/smc\_h1/get\_spectrum.html}. We then determined whether
there was any \hi\ detected in the spectrum, and, if so, the radial velocity of the associated gas. These data are interpreted in \S~\ref{sec:results}.


\begin{table*}
\caption{The clusters and their properties used in the present work, and the results of the \hi\ survey.  The ages and masses have been taken from the compilation of McLaughlin \& van der Marel~(2005) with the exceptions of NGC~1856 and NGC 1866 which are from Bastian \& Silva-Villa~(2013) and NGC~1850 which is from Niederhofer et al.~(2014). }
\label{tab:clusters}
\begin{tabular}{lccccc}
\noalign{\smallskip}
\hline
\hline
\noalign{\smallskip}
Cluster &  Age & Mass  & H{\sc i} & relative velocity & Reference\\
 & Myr & [10$^3$~\msun] &detection& (H{\sc i} \& cluster) & (cluster velocity)\\
\hline
\noalign{\smallskip}
NGC 1711 & 50 & 17 &  no &	$-$ \\
NGC 1818   & 25 &  26&  yes  & $5\pm7$~km/s & Freeman et al. ~1983\\
NGC 1831  & 315  & 39& yes & $>10$~km/s & Schommer et al.~1992	\\
NGC 1847  & 26  & 25& yes & $10$~km/s & this work	\\
NGC 1850  & 100 & 140&  yes & $>10$~km/s & Fischer et al.~1992	\\
NGC 1856  & 280 & 76&  yes & $>10$~km/s  & Freeman et al.~1983	\\
NGC 1866  & 180  & 81&  no &$-$ 	\\
NGC 2031  & 160 & 30&  	yes & $0\pm2$~km/s & Storm et al.~2005\\
NGC 2136   & 100 & 20&  yes & $>10$~km/s & Mucciarelli et al.~2012	\\
NGC 2157   & 40 & 20&  yes & $>10$~km/s	& Freeman et al.~1983\\
NGC 2164  & 50 & 15&  no &	$-$ \\
NGC 2214  & 40 & 11 &  no &	$-$ \\
NGC 330  & 25 & 50& yes  &  $>10$~km/s & Hill~1999\\

\hline
\end{tabular}
\end{table*}

\subsection{Spitzer Imaging}
\label{sec:spitzer}

In order to search for cool/warm dust within the clusters, we used the publicly available {\em Surveying the Agents of Galaxy Evolution} (SAGE) Spitzer Space Telescope survey of the Magellanic Clouds (Meixner et al.~2006). We downloaded the processed images\footnote{http://irsa.ipac.caltech.edu/data/SPITZER/SAGE/ and http://irsa.ipac.caltech.edu/data/SPITZER/SAGE-SMC/} in a cut out around the position of the cluster centres.  Specifically, we used the $3.6$, $70$ and $160$\micron\ images.  NGC~2214 was outside the SAGE field-of-view, so no limits on the dust mass present in that cluster can be given.

\begin{figure}
\includegraphics[width=8.5cm]{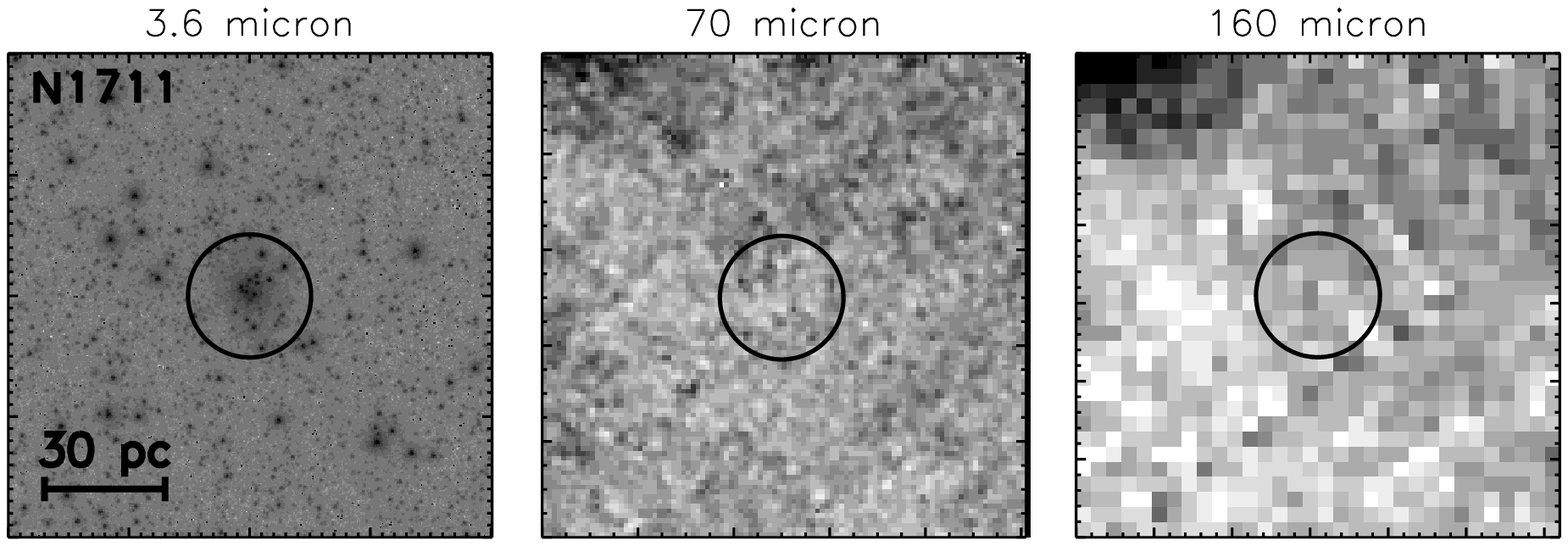}
\includegraphics[width=8.5cm]{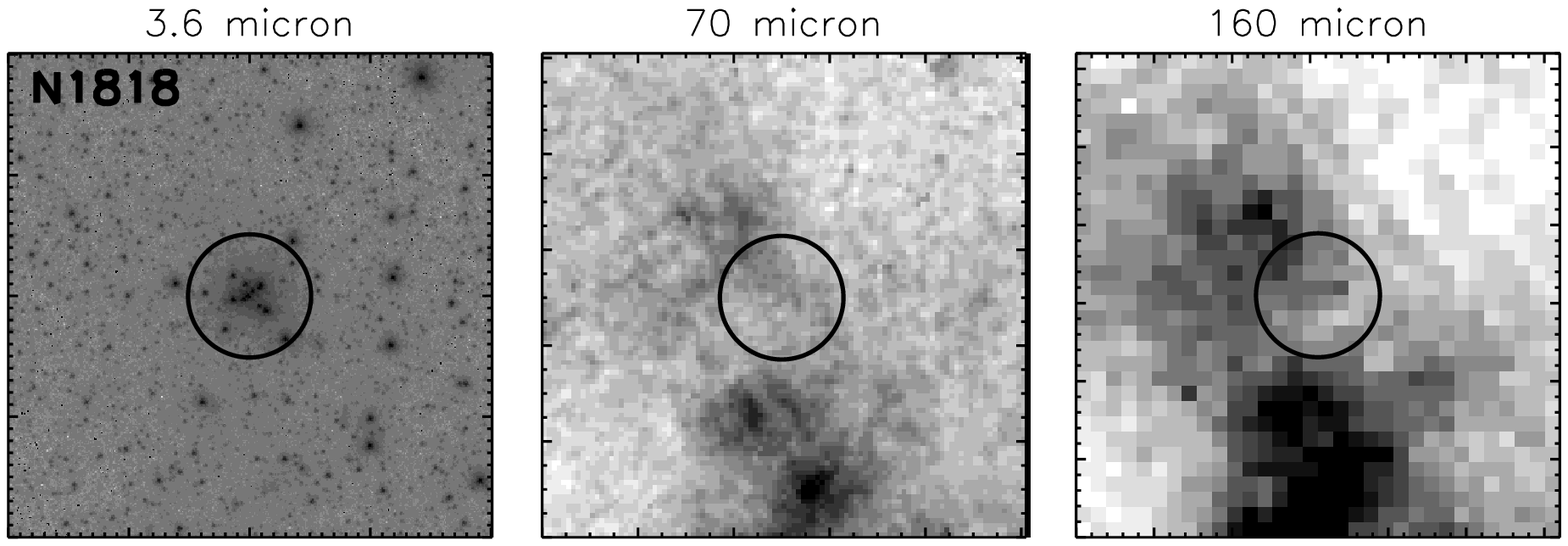}
\includegraphics[width=8.5cm]{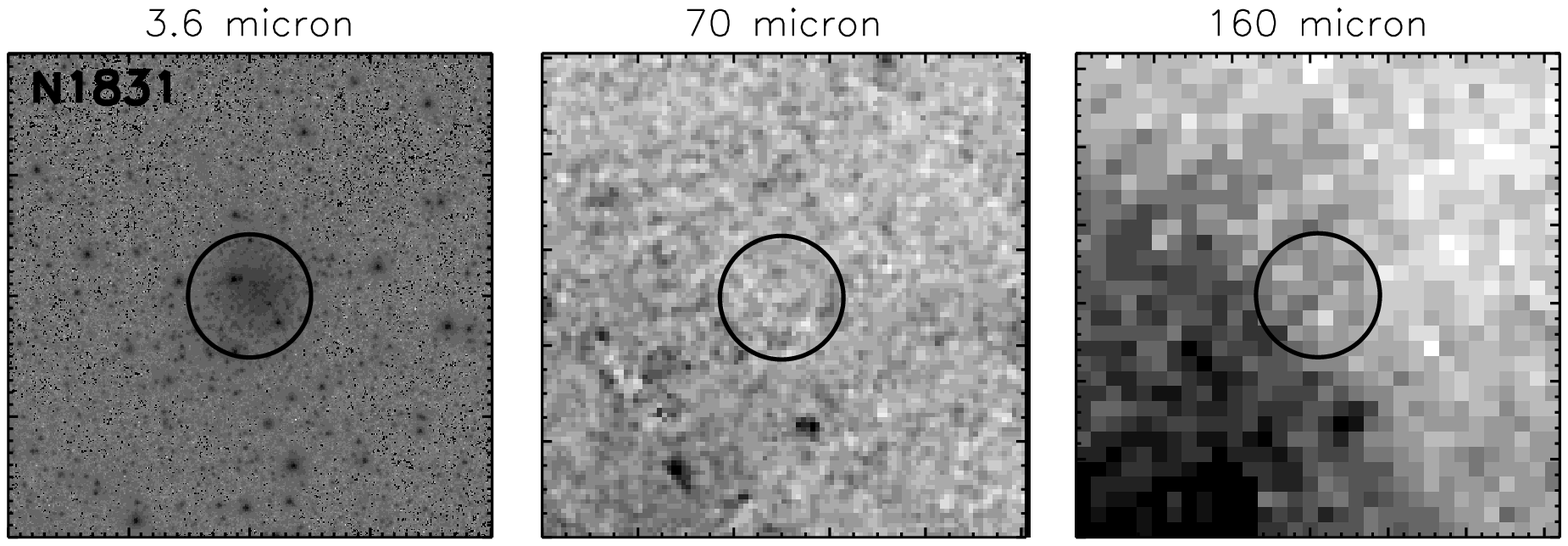}
\includegraphics[width=8.5cm]{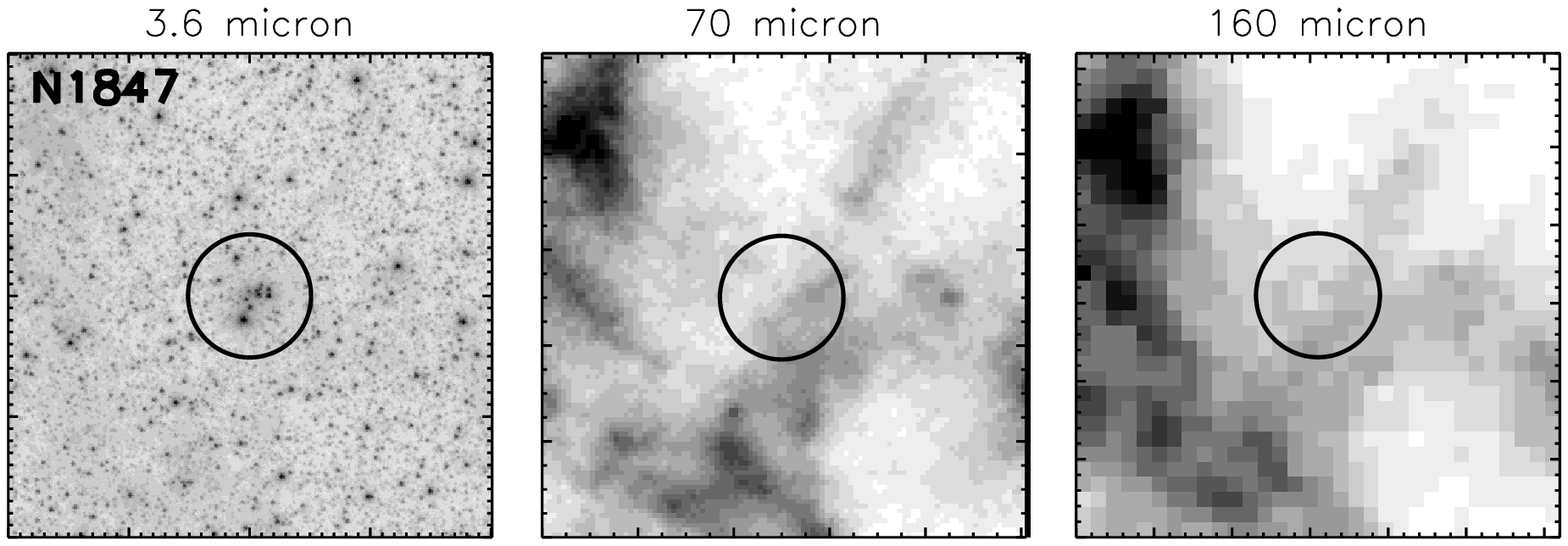}
\includegraphics[width=8.5cm]{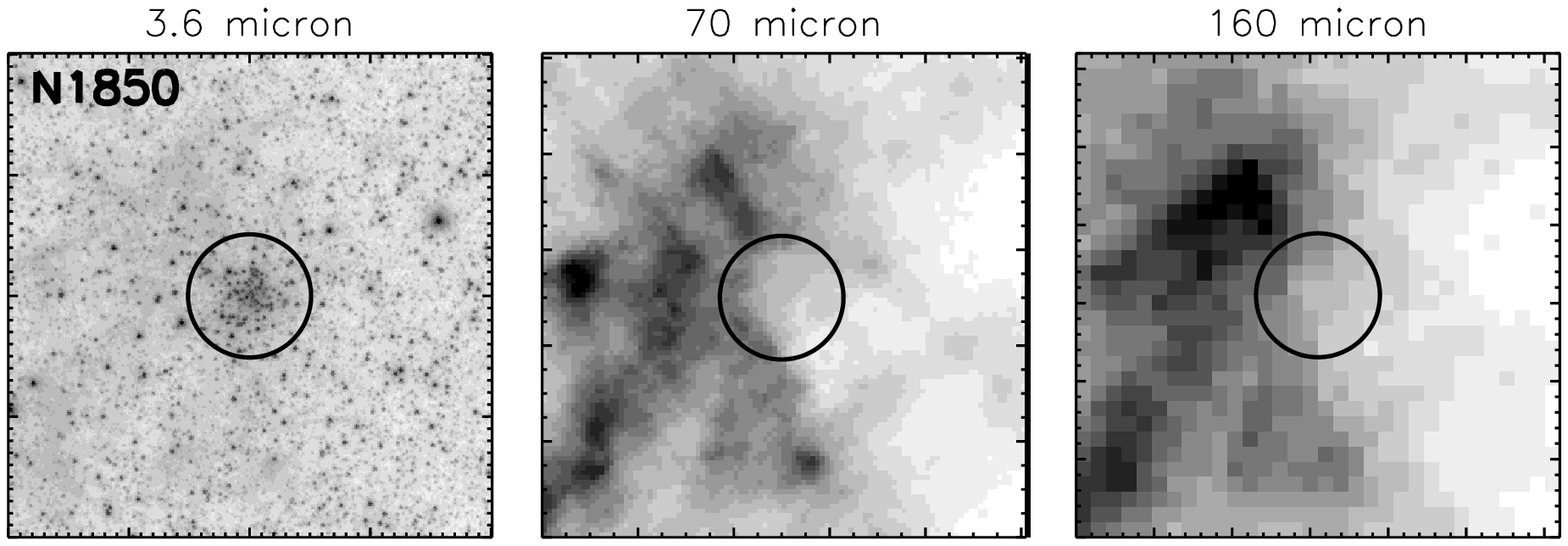}
\includegraphics[width=8.5cm]{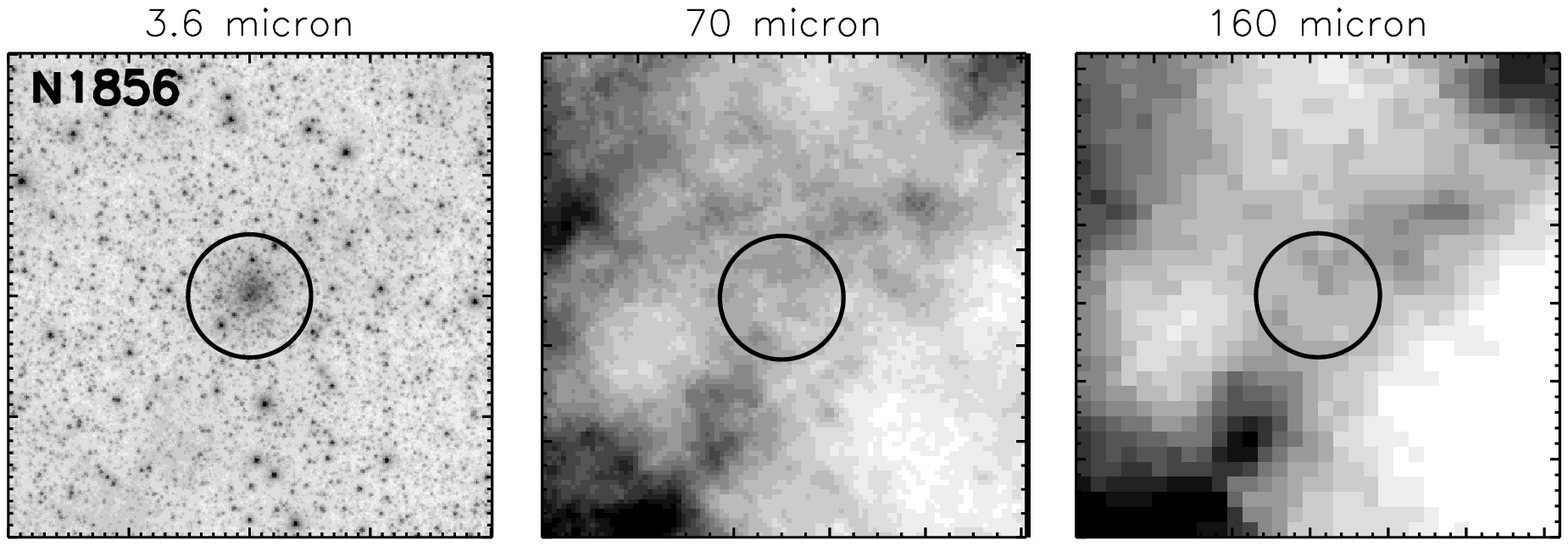}
\includegraphics[width=8.5cm]{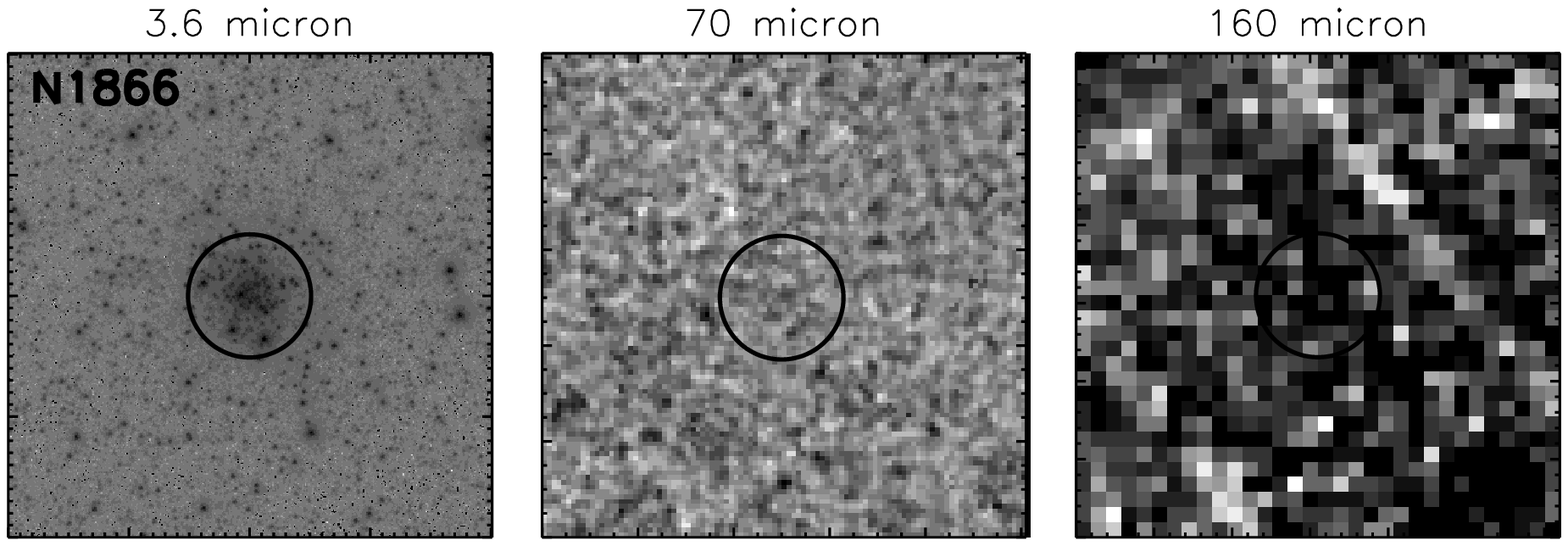}
\caption{Spitzer images of each cluster at $3.6$, $70$, and $160$\micron.  The circle in each image has a radius of 61", which corresponds to $\sim15$~pc and $\sim18$~pc for the LMC and SMC clusters, respectively.  For all the images, dark colours show emission.}
\label{fig:spitzer1}
\end{figure}

\begin{figure}
\includegraphics[width=8.5cm]{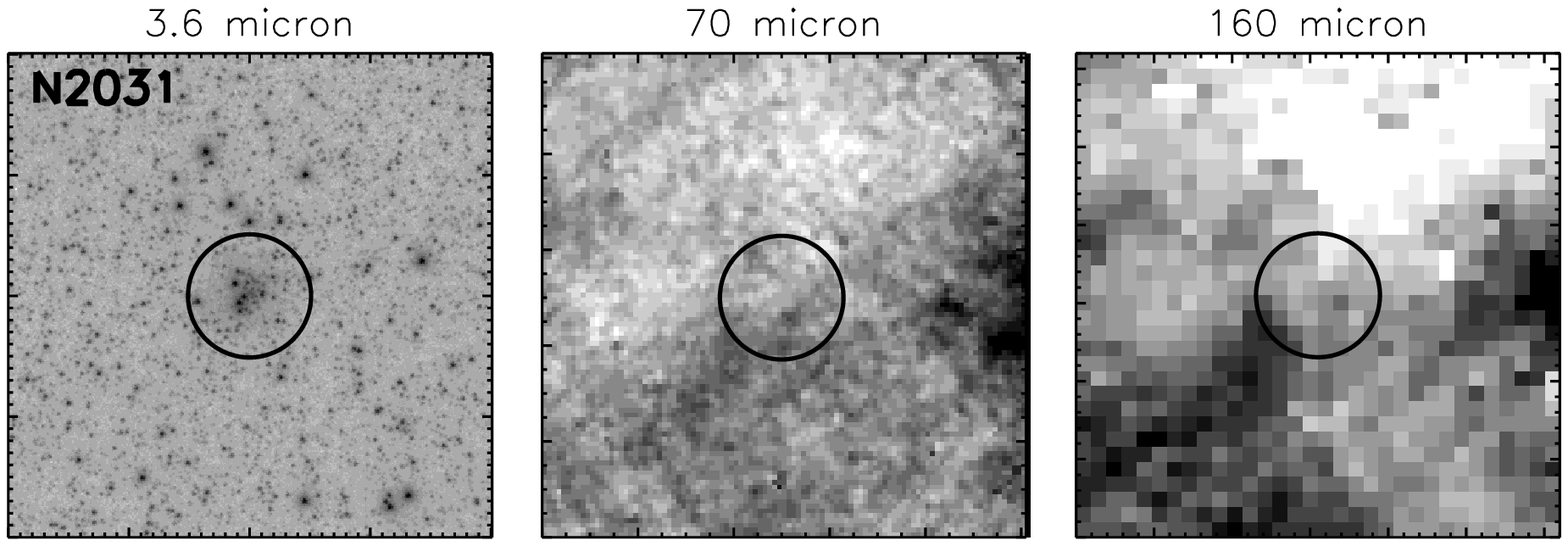}
\includegraphics[width=8.5cm]{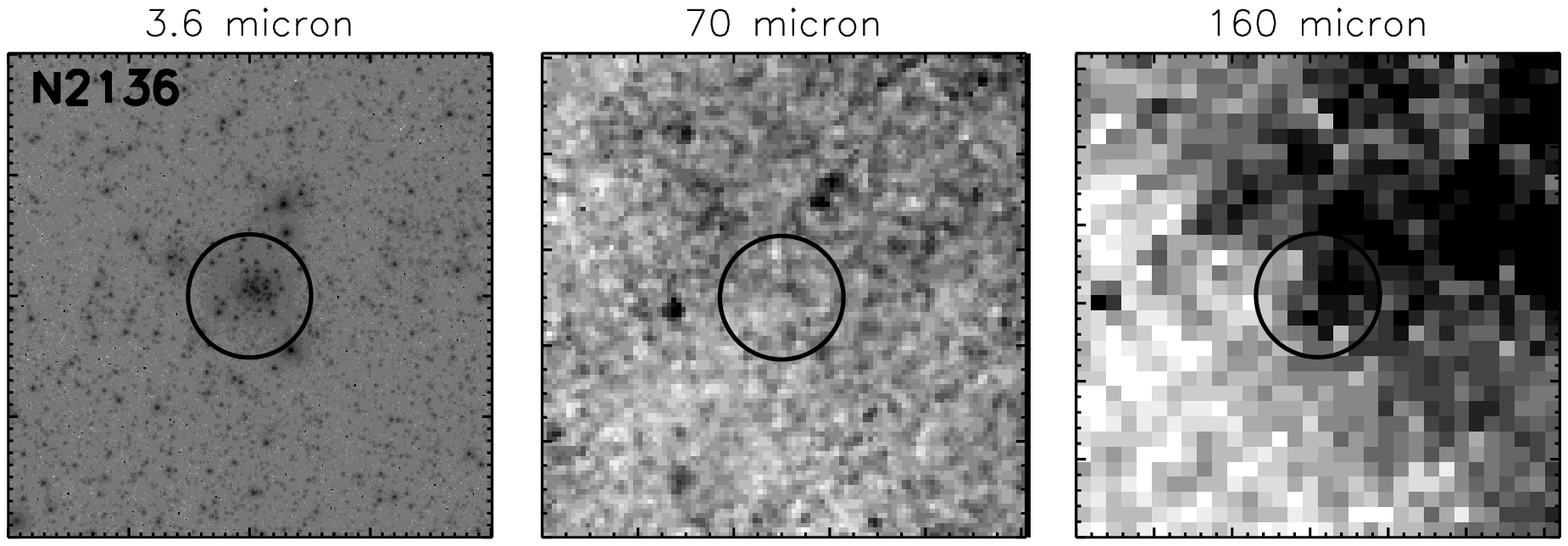}
\includegraphics[width=8.5cm]{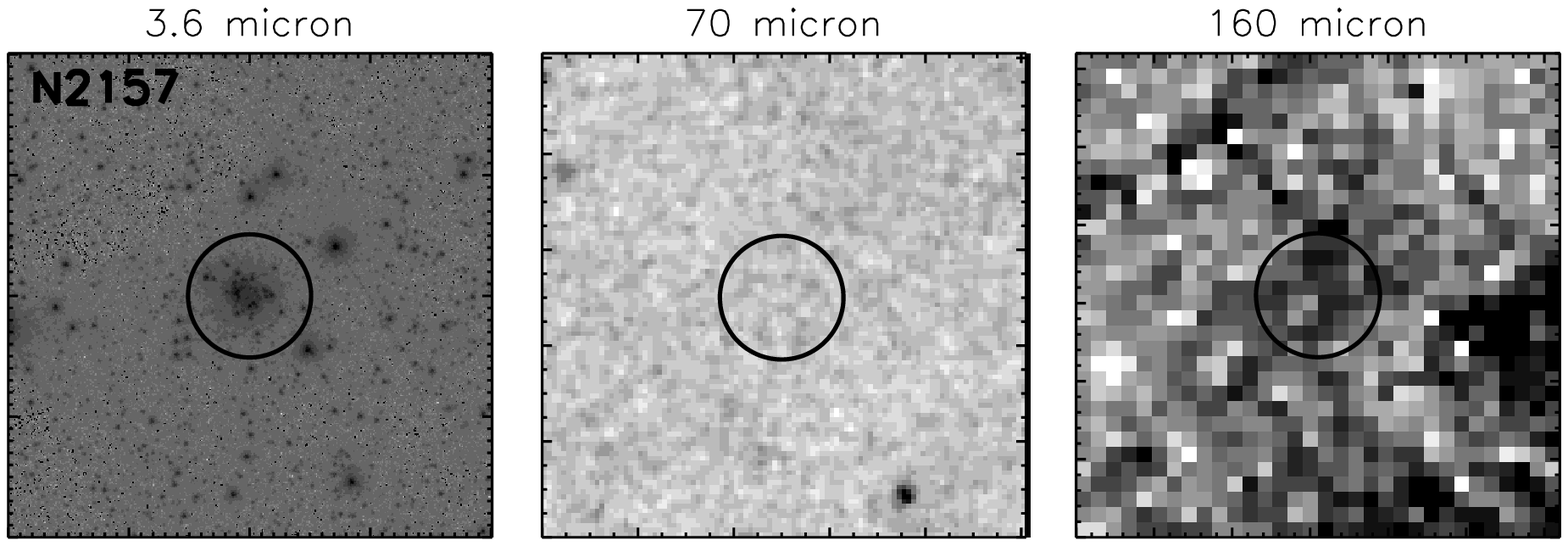}
\includegraphics[width=8.5cm]{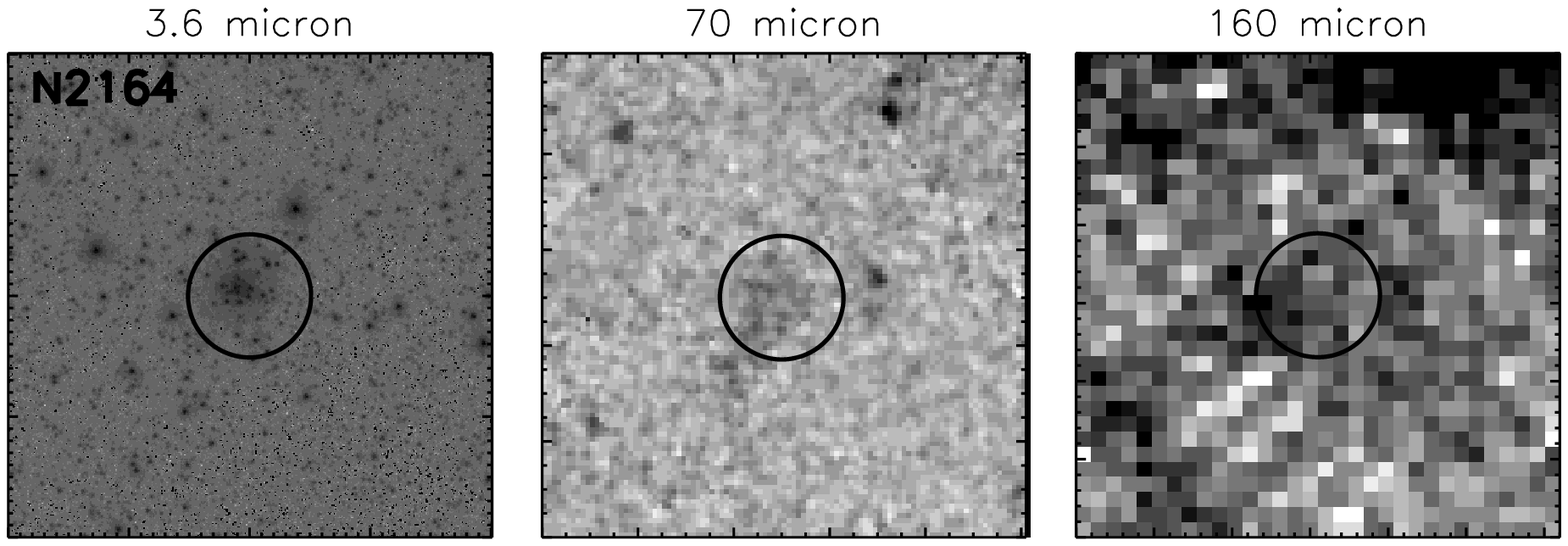}
\includegraphics[width=8.5cm]{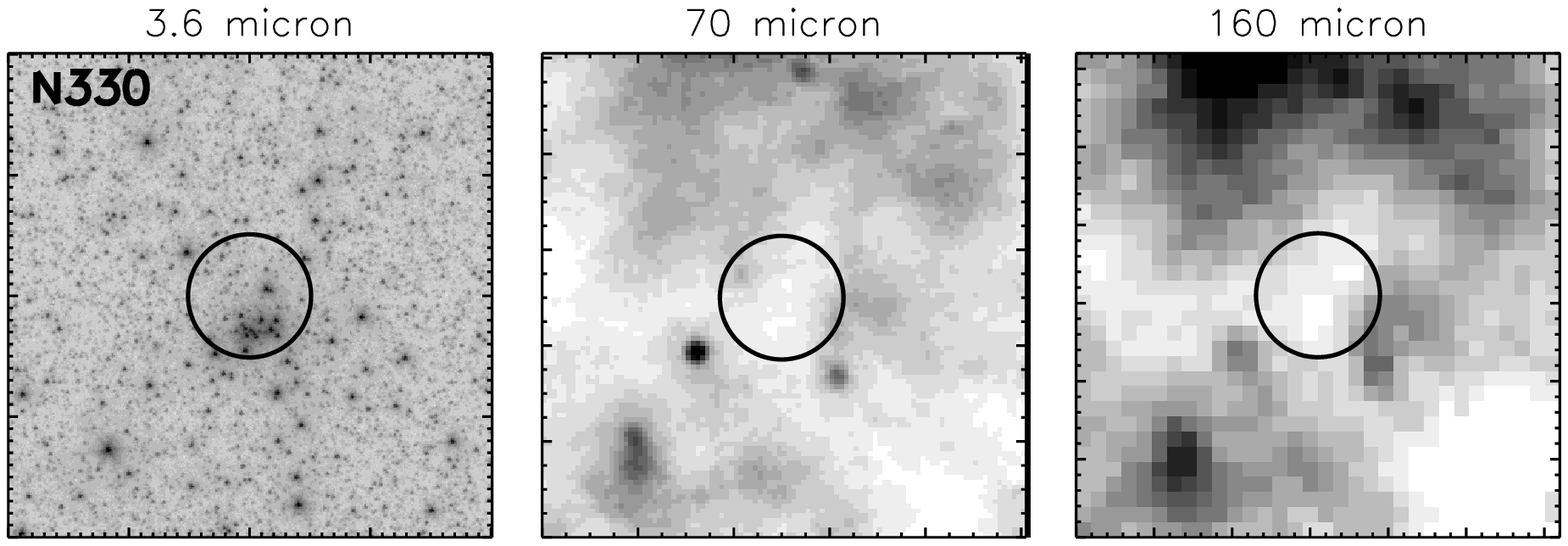}

\caption{Figure~\ref{fig:spitzer1} continuued.}
\label{fig:spitzer2}
\end{figure}

The $3.6$~\micron\ image was used to locate the position of the cluster, as stars with cool photospheres (giants) are readily detected.  The $70$ and $160$~\micron\ images were then used to search for dust within the clusters.   Images of the clusters in the three wavelengths are shown in Figs.~\ref{fig:spitzer1} \& ~\ref{fig:spitzer2}.

With the possible exception of NGC~2136 and 2164 (see \S~\ref{sec:spitzer_sensitivity}), we did not detect emission at $70$ or $160$~\micron\ for any of the clusters.  Like for the \hi, some clusters (e.g., NGC~2031) did have emission within the aperture, however this emission was not centred on the clusters. Instead, it appears to be related to nearby filaments.  NGC~2157 may have a slight detection in the $160$~\micron\ image, although no corresponding emission was found in the 70~\micron\ image.

\subsection{SOAR observations}

One of the LMC clusters in our sample (NGC 1847) had no published radial velocity to compare to the detection of \hi\ at its position. To estimate its velocity, we obtained longslit spectra of two bright stars near the cluster center using the Goodman High-Throughput Spectrograph (Clemens et
al. 2004) on the SOAR 4.1-m telescope. We used a 2100 l/mm grating and a 1.03\arcsec\ slit for the single 300-sec observation, giving a resolution of 0.9 \AA\ over the wavelength range $\sim 4960$--5600 \AA. The spectra were reduced in the standard manner, and heliocentric radial velocities derived through cross-correlation with standard stars observed with the same setup. The two stars had radial velocities consistent to within 3 km/s, suggesting they are both cluster members. The mean velocity is $292\pm3$ km/s, which we adopt as the systemic radial velocity for NGC 1847.

\section{Results}
\label{sec:results}

\subsection{HI sensitivity and detections}
 
The results from the \hi\ survey are presented in Table 1. Eleven of the 13 clusters in our sample did not have \hi\ detected at the position and velocity of the cluster. In order to determine the upper limit to the amount of \hi\ gas present, we generously take the range $\pm10$ km/s from the cluster (20 km/s total). For the typical noise of the LMC and SMC surveys (the SMC survey was somewhat deeper), this gives $3\sigma$ column density limits of $7.9\times10^{19}$ atoms/cm$^{2}$ and $4.3\times10^{19}$ atoms/cm$^{2}$, respectively. To convert these values into \hi\ masses, we need to assume a size of the cluster, and find $M_{\hi,LMC} = 1.99 R_{\rm cl}^2 M_{\odot}$ (for the single SMC cluster in our sample, the prefactor is 1.08 rather than 1.99). For a radius of 10 pc, the $3\sigma$ upper limit for the LMC clusters is 199 $M_{\odot}$. If we were less conservative and adopted $\pm5$ km/s, this value would be lower by a factor of 2. Similarly, if the gas was more concentrated, i.e., within a radius of 5 pc, the upper limit would shift to $\sim 50 M_{\odot}$.

\begin{figure}
\includegraphics[width=8.5cm]{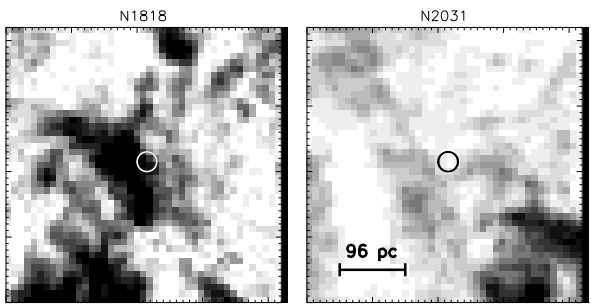}
\caption{\hi\ emission maps for the two clusters (NGC~1818 and NGC~2013) that have detected \hi\ within an aperture of 15~pc of the cluster with the same (within the errors) radial velocity.  The circles show a radius of 15~pc (the same as used in Fig.~\ref{fig:spitzer1}).  Note that in both clusters the emission does not appear to be from the clusters, but rather due nearby larger scale filaments.}
\label{fig:hi_images}
\end{figure}

Two of the clusters (NGC 1818 and NGC 2031) have \hi\ detected at the spatial position (with the 15 pc aperture) and velocity of the cluster. The inferred \hi\ mass summed over a range $\pm10$ km/s around the peak velocity along the line of sight for NGC 1818 is $M_{\hi} = 26 \times R_{\rm cl}^2~ M_{\odot}$, i.e., $\sim 2600~M_{\odot}$ for a radius of 10 pc. However, spatially, the gas is not centred on the cluster, but rather appears to be projected on the edge of a large \hi\ cloud to its northeast. This is shown in Fig.~\ref{fig:hi_images}. Hence, it is reasonable to argue that the \hi\ is not likely to be associated with the cluster. However, we cannot prove or disprove this with the currently available \hi\ data alone; better resolution imaging is required. However, looking at the Spitzer images (see \S 3.2), which have higher resolution (Fig.~1), we see that the emission is not coming from the cluster itself, but rather from nearby filamentary gas/dust clouds.

For NGC 2031, the inferred \hi\ mass is $M_{\hi} = 17 \times R_{\rm cl}^2~ M_{\odot}$, corresponding to $\sim 1700 M_{\odot}$ for a radius of 10 pc. However, NGC 2031 is located (in projection) on the edge of a filament (also seen in cold dust in Fig. 2), suggesting that the detected \hi\ is not associated with the cluster directly. As was the case for NGC 1818, higher resolution data would be required to assess whether the \hi\ was likely associated with the cluster. Again, the Spitzer imaging shows (Fig.~2) that the emission is not due to the cluster, but rather from nearby filamentary clouds. Hence, from the current data, it appears that the gas is not physically associated with the clusters.

\subsection{Spitzer sensitivity}
\label{sec:spitzer_sensitivity}

In order to determine the sensitivity of the observations we use the approach of Gordon et al.~(2010 - their Eq.~2) to estimate the amount of dust mass present in both the $70$ and $160$~\micron\ images.  Here the dust mass is a function of the grain size, density, temperature, distance, flux, and the emissivity of the dust at that wavelength.  We adopt the same parameters as Gordon et al.~(2010), and an emissivity of the dust at $70$~\micron\ that is five times higher than at $160$~\micron.   We note that if we used the more recent emissivity value at $70$~\micron\ of 3.4 times higher $160$~\micron\ (Planck collaboration, Ade et al.~2014), the estimated masses at $70$~\micron\ would be increased by a factor of $\sim1.5$, which would not significantly impact our results.

\begin{figure}
\includegraphics[width=8.5cm]{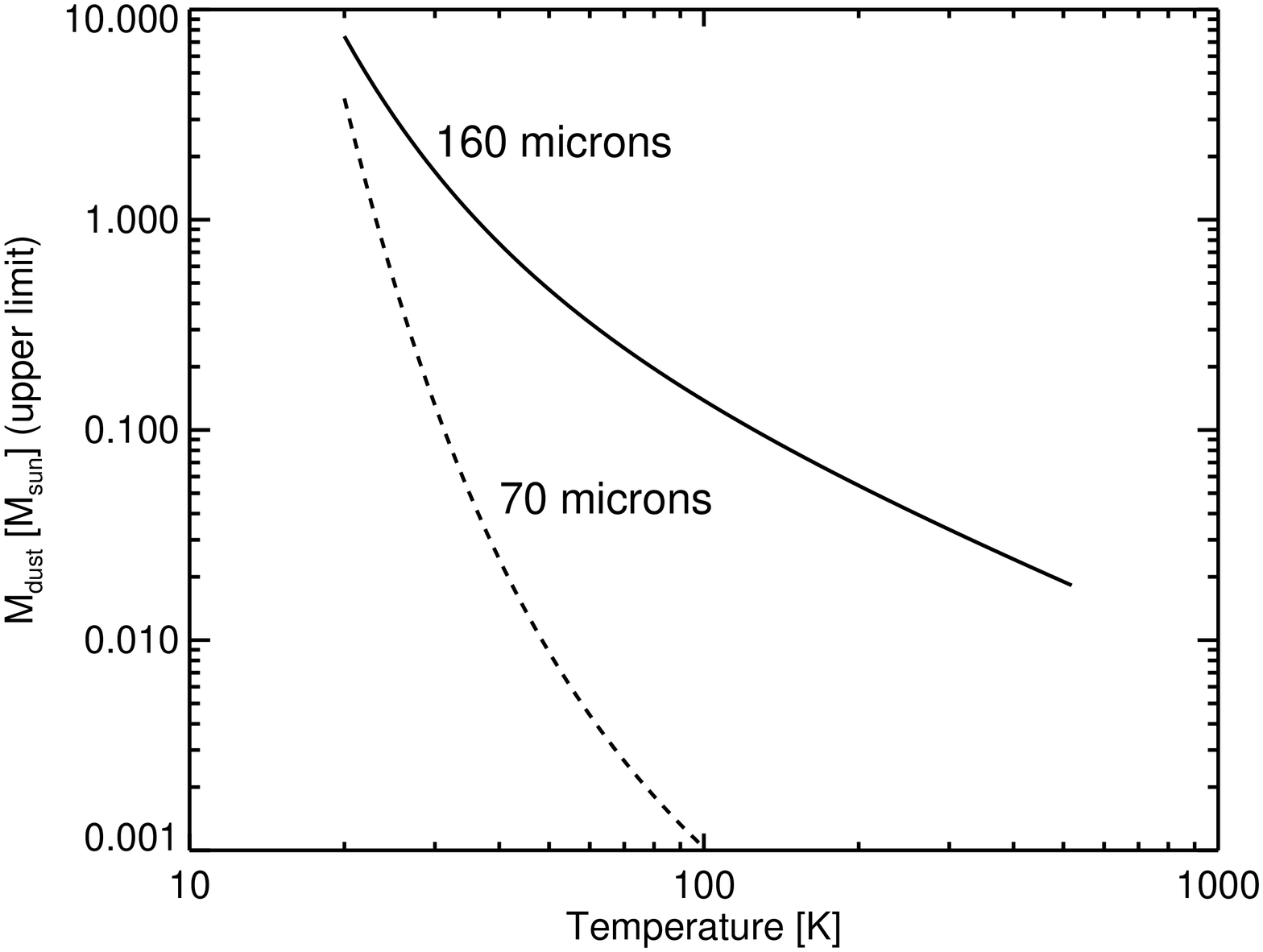}
\caption{The estimated upper limit on the amount of dust present within the clusters, based on the detection limit for each of the Spitzer bands used.}
\label{fig:dust}
\end{figure}

Since most of the clusters had a clear detection at the location of the clusters, we adopt the flux limits given by the {\em SAGE} survey, namely $0.1$ and $1$~Jy at $70$ and $160$~\micron, respectively.  This just leaves the dust temperature as a free parameter.  Adopting these limits, the upper limit to the amount of dust present (in each of the bands) as a function of temperature is shown in Fig.~\ref{fig:dust}.  Note the strong temperature dependence on the upper limit of the dust mass.  

If we adopt a dust temperature of $30$~K, we obtain an upper limit of $\sim0.15$ and $\sim2$~\msun\ for the dust mass for the $70$ and $160$\micron\ fluxes, respectively.  Assuming a gas to dust ratio of 500-to-1 (e.g., van Loon et al.~2005; Roman-Duval et al.~2010), this leads to an upper limit of $75$ and $1000$~\msun\ of gas for the two bands.  However, if the dust was this cool, we may expect high extinction in the optical, or for the gas to catastrophically cool (CS11) and begin forming stars.  Hence, if gas does exist within the clusters, we may expect it to be at a higher temperature.  Adopting a dust temperature of $100$~K leads to significantly lower upper limits on the dust mass, $0.001$ and $0.15$~\msun, for the $70$ and $160$\micron\ fluxes, respectively.  Again, assuming a gas-to-dust ratio of 500-to-1, leads to upper limits on the total gas mass within any of the clusters of $0.5$ and $75$~\msun\ for the two bands, respectively.

Due to the steep dependence of the upper limit of the dust mass on the dust temperature, temperatures above $100$~K lead to dust and gas masses well below 1\% of the stellar mass of the cluster.

NGC~2136 has a significant amount of flux in the $160$\micron\ image (see Fig.~\ref{fig:spitzer2}), which appears to be an extension of a nearby gas/dust cloud.  No detection is found in the corresponding $70$\micron\ image.

\subsection{The curious case of NGC~2164}
\label{sec:n2164}
 
One cluster, NGC~2164, shows a marginal detection at 70$\micron$. While most of the emission is not centred on the cluster, the cluster resides at one end of the extended emission, suggestive that the emission may originate from the cluster. Less than 10\% of the $70\micron$ flux of the extended feature is seen within 15~pc of the cluster centre. No trace of the extended emission is detected in the $24\micron$ or $160\micron$ images.  Additionally,  there is no \hi\ emission detected at any velocity at the position of the cluster (or along the extended filament).

In order to see if there is indeed dust within the cluster, we used archival HST F555W (V) and F814W (I) band imaging in order to search for a colour gradient within the cluster.  If there is gas/dust within the cluster, the expectation is that it would collect in the centre of the cluster and be more centrally concentrated than the stars (e.g., CS11).  As such, we would expect the integrated (radially) cluster colour to become redder closer to the centre (e.g., Bastian \& Silva-Villa~2013).  We do not find any colour gradient within the cluster from $1-5$~pc from the cluster centre.  Within $1$~pc, however, the colour becomes bluer at smaller radii.  The overall, $<5$~pc, colour of the cluster (F555W--F814W $= 0.6$) is consistent with an age of $\sim40$~Myr (McLaughlin \& van der Marel~2005) and a small amount of extinction, $A_{V} \lesssim 0.3$~mag (e.g., Kotulla et al.~2009).  Hence, we also do not find evidence for dust within the cluster based on optical imaging.

The origin of the extended $70\micron$ emission near NGC~2164, with no counterpart detected at any other wavelength, remains unknown.

\subsection{Dust production within the clusters}
\label{sec:dust_clusters}

While no \hi\ gas or dust was observed within the clusters studied here, many of the clusters studied here have luminous evolved stars that are expected to produce dust.   Would we expect to observe such dust?  The average cluster in our sample is expected to have ten or fewer luminous AGB stars (Mucciarelli et al.~2006), whereas the younger clusters (e.g., NGC~1818, NGC~1847 and NGC~330) may have up to $\sim20-30$ red super-giants (RSGs) within them (e.g., Davies et al. 2008).  The typical total (gas + dust) mass-loss rates for AGB and RSGs in the LMC are $\sim1\times10^{-6}$~\msun/yr (Bonanos et al.~2010).

Assuming a gas-to-dust ratio of 500-to-1, the amount of dust produced within the clusters per Myr is expected to be $\sim0.04-0.06$~\msun\ or less.  If the temperature of the dust is larger than $\sim30$~K, it may be detectable at 70~\micron\ (see Fig.~\ref{fig:dust}).  The fact that we do not detect the clusters at 70~\micron\ (with the possible exception of NGC~2164) means that either the assumed number of evolved massive stars (and their mass-loss rates) are overestimated (as they are meant to be upper limits), or the clusters are efficient at removing any gas/dust released in the cluster.  We note that some of the younger clusters (e.g., NGC~330) may have a significant number of Be stars within them that may also contribute gas and dust to the host cluster (e.g., Martayan et 2007), further emphasising the efficiency of the gas/dust removal process from young ($\lesssim300$~Myr) clusters.  Also, we note that the mass loss rates of RSGs and AGBs are expected to be highly time variable, with the high mass loss rate phase being short-lived (e.g., van Loon, Marshall, \& Zijlstra~2005).  Hence, in order to set strict limits on the amount of gas/dust expected, detailed study of the stellar populations within each cluster is required, meaning that the above calculations are meant to be indicative only.


The winds of AGB stars in the LMC have velocities between $8$ and $25$~km/s (e.g., Marshall et al.~2004), which are at, or near, the escape velocities of the clusters in our sample (Niederhofer et al. 2014), hence it is possible that the RSG/AGB ejecta are able to flow freely from the cluster.  Additionally, radiation pressure and the ionising flux from stars still on the main sequence (or near the main sequence turnoff) will also act to remove gas/dust from the clusters.  Finally, since the amount of gas/dust from RSG/AGB stars within the cluster, at any given time, is expected to be a small fraction of the cluster mass $<1$\%, it is possible that the gas/dust is efficiently stripped by the intra-cluster medium as the cluster moves through the galaxy (CS11).  In a future work, we will place constraints on the amount of dust within much more massive (although more distant) clusters than those presented here.

\section{Discussion and Conclusions}
\label{sec:discussion}

We have searched for evidence of \hi\ gas and/or dust within a sample of massive ($>10^4$~\msun) clusters within the Large and Small Magellanic Clouds.  The clusters have ages between $\sim25$ and $\sim300$~Myr, hence would be expected to contain gas if they were able to form a second (or continuous) generation of stars within them, as predicted by recent theory (e.g., CS11).  In the majority of our sample, no gas or dust was found at the location (and velocity) of the clusters.  In two cases, \hi\ gas was detected at the location and velocity of the cluster (NGC~1818 and NGC~2031), however, upon further examination, the gas was not found to be centred on the clusters, but rather appears to belong to nearby filaments.  

\begin{figure}
\includegraphics[width=8.5cm]{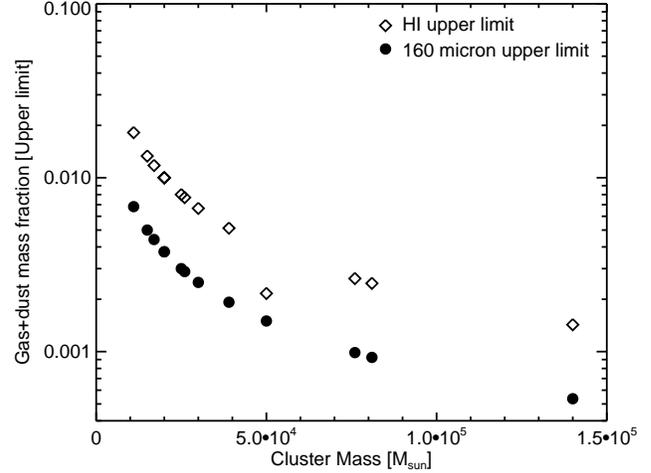}
\caption{The estimated upper limit on the mass fraction of gas (and dust) to stars within the clusters.  The limits from \hi\ and $160\micron$ are shown, and the limits from the $70\micron$ are well below those shown. }
\label{fig:mass_fraction}
\end{figure}

We have placed upper limits of $\sim200$~\msun\ of gas within the clusters based on $\hi$ measurements and $\sim1-75$~\msun\ of total gas based on the dust measurements.  In Fig.~\ref{fig:mass_fraction} we show the upper limit on the fraction of gas+dust within the cluster compared to the stellar mass as a function of stellar mass.  In all cases, we can place strict upper limits of $<10$\% of the (stellar) mass of clusters is present in gas+dust, and in most cases the upper limit is $<1$\%.  Hence, we conclude that these clusters are not able to retain or accrete significant amounts of gas.

Our observations are in conflict with the predictions of CS11, who predicted that all clusters in the LMC with masses above $10^4$~\msun (and have ages $\gtrsim30$~Myr) should be able to retain ejected gas from the stars within them and accrete new gas from their surroundings.  They use this prediction to explain the potential existence of large ($200-500$~Myr) age spreads in intermediate age clusters in the LMC (e.g., Goudfrooij et al.~2011b).  Additionally, they use their predictions to support the ``AGB scenario" for the formation of multiple populations within clusters, where globular clusters must retain the AGB ejecta of stars within them and accrete large amounts of ``pristine" material in order to form a second generation of stars within the young cluster.

The lack of observed gas, in contradiction to the above prediction, is likely to be due to one of two reasons (or a combination of both):

1) stellar feedback within cluster is not negligible, causing gas to be efficiently removed from the cluster.

2) the accretion of gas from the surroundings is less efficient than assumed.  CS11 assume that $10$\% of the mass of the cluster is already initially in gas within the cluster.  This gas acts as a net that can sweep up additional material from the ISM.  If the simulations start with no gas in the cluster, Bondi-Hoyle accretion is not enough to accrete gas to the necessary amounts to sweep up further material.  The initial gas has to come from somewhere in order to initiate the gas accretion (i.e., the sweep up).  If no gas is initially available then not enough further gas will be accreted by the cluster.

For comparison, NGC~1850 at an age of $\sim100$~Myr, has a stellar mass of $\sim1.4 \times 10^5$~\msun, hence, the assumption is that it should have $\sim14000$~\msun\ of gas within it.  The upper limits imposed by the observations presented here are $\sim1 - 200$~\msun\ (depending on the gas/dust temperature and whether one adopts the limits from \hi\ or dust).  The same can be said about the older massive clusters, NGC~1866 and NGC~1856 (180 and 280~Myr, respectively), both would be expected to have more than $7000$~\msun\ worth of gas within them, or more if they were able to accrete new gas.  The lack of gas within young clusters likely explains why these massive clusters do not show evidence for age spreads within them (Bastian \& Silva-Villa~2013; Niederhofer et al.~2014).

 In the CS11 model, the collected material (i.e. that shed from the stars and accreted from the surroundings) may not be neutral, but rather reside in dense clumps that are ionised by the large number of B-stars in the clusters.  This has not been tested with the present observations, however such ionised clumps would be expected to produce strong optical emission lines.  In a survey of 130 Galactic and extragalactic massive clusters ($10^4 < M/\msun < 10^8$) with ages greater than 10~Myr, none were found with emission lines in their integrated spectrum, implying that there is $<100$~\msun\ of ionised gas within the clusters.  Additionally, optical spectroscopy of LMC clusters has not detected emission lines associated with the clusters (e.g., Beasley et al. 2002;  Leonardi \& Rose~2003; Palma et al. 2008).

Whether more massive clusters are able to retain and accrete significant amounts of gas remains unclear.  Massive old globular clusters display a well known deficit of gas within them, relative to the observed outflows of their red giant stars within them, showing that even massive clusters (with low stellar feedback relative to that expected in young clusters) are unable to hold on to gas within them (e.g., van Loon et al.~2009 and references therein).  Potentially, observations with {\em ALMA} will be able to find or rule out gas in more distant young clusters with masses in excess of $10^5$ or $10^6$~\msun.  However, we note that no clusters with (main population) ages above 10~Myr, including those with masses up to $10^8$\msun, have been found with evidence of ongoing star-formation within them (Bastian et al.~2013a).  Additionally, integrated spectroscopy of a young massive cluster ($\sim10^7$~\msun) has shown that it is consistent with single burst of star-formation (Cabrera-Ziri et al.~2014).  

Many popular models for the production of multiple generations of stars in massive clusters predict that significant neutral gas must be present in the clusters for the second generation(s) to form. Observations presented in this paper, especially when combined with the above results, appear to be in strong tension with these predictions.

\section*{Acknowledgments}

We would like to thank Mikako Matsuura for helpful discussions on dust masses and the SAGE survey and Charlie Conroy, Laura Chomiuk and Steve Longmore for helpful comments and discussions. We would also like to thank  Alex McComb for help with data reduction.  The referee, Jacco van Loon, is gratefully acknowledged for his helpful and constructive report.  NB is partially funded by a Royal Society University Research Fellowship.  This paper was partially based on observations obtained at the Southern Astrophysical Research (SOAR) telescope, which is a joint project of the MinistŽrio da Cincia, Tecnologia, e Inova‹o (MCTI) da Repœblica Federativa do Brasil, the U.S. National Optical Astronomy Observatory (NOAO), the University of North Carolina at Chapel Hill (UNC), and Michigan State University (MSU).

\bsp
\label{lastpage}
\end{document}